# PHP code smells in web apps: survival and anomalies

Américo Rio[1,2,*] and Fernando Brito e Abreu[1]

[1] Iscte - Instituto Universitário de Lisboa, ISTAR, Lisboa, Portugal
`{jaasr,fba}@iscte-iul.pt`
[2] NOVAIMS, UNL, Lisboa, Portugal
`americo.rio@novaims.unl.pt`
*corresponding author*

**Abstract.**

**Context**: Code smells are considered symptoms of poor design, leading to future problems, such as reduced maintainability. Except for anecdotal cases (e. g. code dropout), a code smell survives until it gets explicitly refactored or removed. This paper presents a longitudinal study on the survival of code smells for web apps built with PHP, the most widely used server-side programming language in web development.

**Objectives**: We aim at answering four research questions: (i) code smells' survival depends on their scope? (ii) practitioners' attitudes towards code smells' removal in web apps have changed throughout time? (iii) how long code smells survive in web applications? (iv) are there sudden variations (anomalies) in the density of code smells through the evolution of web apps?

**Method**: We analyze the evolution of 6 code smells in 8 web applications written in PHP at the server side, across several years, using the survival analysis technique. We classify code smells according to scope in two categories: scattered and localized. Scattered code smells are expected to be more harmful since their influence is not circumscribed as in localized code smells. We split the observations for each web app into two equal and consecutive timeframes, spanning several years, to test the hypothesis that code smells awareness has increased throughout time. As for the anomalies, we standardize their detection criteria.

**Results**: We present some evidence that code smells survival depends on their scope. A trend is not observed in code smells removal across all apps: the average survival rate decreases in some of them, while the opposite is observed for the remainder. Possible interpretations of these different behaviors are put forward. The survival of localized code smells is around 4 years, while the scattered ones live around 5 years. Around 60% of the smells are removed, and some live through all the application life. We also show how a graphical representation of anomalies found in the evolution of code smells allows unveiling the story of a development project and make managers aware of the need for enforcing regular refactoring practices.

**Keywords**: code smells, PHP, software evolution, survival analysis, web apps.

## 1 Introduction

### 1.1 Motivation

Code smells (CS) are symptoms of poor design and implementation choices, therefore fostering problems like increased defect incidence, poorer code comprehension, and longer times to release, in studies involving desktop applications, being Java the most studied language in this aspect (Rasool & Arshad, 2015), (Singh & Kaur, 2018). Code smell is a term introduced by Kent Beck, in a chapter of the famous Martin Fowler's book (Fowler, 1999), to describe a surface indication that usually corresponds to a deeper problem in the system. A smell is something relatively quick to spot, like a long method, or many parameters in a method. Although smells do not always indicate a problem, often they do. Keeping CS in the code may be considered a manifestation of *technical debt*, a term coined by Ward Cunningham almost three decades ago (Cunningham, 1992), a metaphor for the trade-off between writing clean code at a higher cost and delayed delivery, and writing messy code cheap and fast at the cost of higher maintenance efforts (Buschmann, 2011).

The Software Engineering community has been proposing new techniques and tools, both for CS detection and refactoring (Fernandes et al., 2016; Reis et al., 2020; Zhang et al., 2011), in the expectation that developers become aware of their presence and get rid of them. A good indicator of the success of that quest is the reduction of *CS survival*. The latter is its lifetime, that is, the elapsed time since one CS is introduced until it disappears, either due to refactoring, or to code dropout. Software evolution (longitudinal) studies are required to assess CS survival. Software Evolution is an active research thread in Software Engineering, where longitudinal studies have been conducted on software products or processes, focusing on aspects such as software metrics, teams' activity, defects identification and correction, or time to release (Herraiz et al., 2013; Radjenović et al., 2013).

The work reported in this paper is in the crossroads of Software Evolution (Madhavji et al., 2006; Rudall, 2000) and Web Engineering (Mendes & Mosley, 2006; Rossi et al., 2007) and is a continuation of our research on the evolution of web systems/applications, regarding quality aspects of maintainability and reliability (Americo Rio & Brito e Abreu, 2017; Américo Rio & Brito e Abreu, 2016). Web applications (web apps, for short) are different from desktop applications and mobile applications, since they run both on a browser and a server, and encompass a heterogeneity of target platforms, programming, and content formatting languages. Due to this diversity, it is necessary to performs similar and different studies regarding their specificity.

This paper addresses the survival of CS in the server-side of web apps using the PHP programming language, currently reported to hold 79% of the market share in that sector (W3techs.com, n.d.). It is an extended and updated version of (Américo Rio & Brito e Abreu, 2019). We doubled the number of PHP web apps (from 4 to 8) and performed a more thorough data analysis, including more metrics collection and considering more factors in the data analysis. In this longitudinal study we considered as many years as possible for each web app, summing up to 441 versions. Furthermore, the topic detection of anomalies in CS evolution was not included in the original paper.

The number of evolution studies in web apps, either cross-sectional or longitudinal, especially with PHP, is still small (Amanatidis & Chatzigeorgiou, 2016; Kyriakakis & Chatzigeorgiou, 2014) so we expect the work presented herein may contribute to this body of knowledge, namely by confirming or refuting findings on the survival of CS in the existing literature.

**1.2    Research questions**

CS's effect can vary widely in breadth. In localized ones, the scope is a method or a class (e.g. *Long Method, Long Parameter List, God Class*), while the influence of others may be scattered across large portions of a software system (e.g. *Shotgun Surgery, Deep Inheritance Hierarchies*, or *Coupling Between Classes*). Since widespread CS can cause more damage than localized ones, a possible behavior is that practitioners assign a higher priority to their removal and, as such, their survival rates are shorter than for localized CS. However, if the current removal of CS is not a priority, there is no time, or the team is small, the contrary can happen, especially because scattered CS require more costly refactorings, especially if done manually. In this paper we aim at providing an answer to the following research question in the context of web apps using PHP as server language:

***RQ1*** *– The survival of localized CS is the same as for scattered CS?*

The other factor we are concerned with regards to a superordinate temporal analysis. Since the topic of CS has been addressed by researchers, taught at universities, and discussed by practitioners[1] over the last two decades, we want to investigate whether this had an impact on CS survival. We expect that, in a long term, increased awareness has caused a more proactive attitude towards CS detection and removal (through refactoring actions), thus leading to shorter survival rates. Summing up, we aim at answering the following research question in the context of web apps using PHP as the server language:

***RQ2*** *– The survival of CS has varied over time?*

It is also important to characterize quantitatively the survival of CS, i.e., how long do they last before they are removed, and the removal rate. Thus, we must answer the following question in the context of web apps using PHP as the server language:

***RQ3*** *– What is the survival of CS?*

During a preliminary data collection, we noticed that in some versions of the target web apps it appears to be some sudden changes in the density of CS[2]. These anomalous situations, that may occur in both directions (steep increase or steep decrease), deserve our attention, either for recovering the story of a project or, if used just-in-time (e.g., integrated into a continuous delivery pipeline), to provide awareness to decision-makers that something unusual is taking place for good or bad. Therefore, we pose yet another research question, also in the context of web apps using PHP as the server language:

***RQ4*** *– Is it possible to detect anomalous situations in the evolution of CS?*

To answer the research questions, we performed a longitudinal study encompassing 8 web apps, and 6 CS, as surrogates of more scattered or localized scopes.

This paper is structured as follows: section 2 introduces the study design; section 3 describes the results of our data analysis, including the identification of validity threats; section 4 overviews the related work on longitudinal studies on CS and in web apps; and section 5 outlines the major conclusions and outlines required future work.

## 2    Longitudinal study design

### 2.1    Applications sample

The aim of this work is to study the evolution and survival of CS in web apps built with PHP in the server-side. The inclusion and exclusion criteria used for selecting the sample of PHP web apps were the following:

---

[1] - https://stackoverflow.com/search?q=code+smells
[2] - Number of code smells divided by a code size metric

Inclusion criteria:

    (i)       the code should be available (i.e., should be open source)
    (ii)      complete applications / self-contained applications, taken from the *GitHub* top listings
    (iii)     programmed with an object-oriented style (OOP)[3]
    (iv)     covering a long period of time (minimum 5 years, more if possible)

Exclusion criteria:

    (i)       libraries
    (ii)      frameworks or applications used to build other applications
    (iii)     web apps built using a framework

For the first study we considered 4 applications with this criterion. As we extended the previous study, we considered the same criterion, with the same application domains[4], although doubling the number of apps per domain. We excluded frameworks and libraries because we want to study typical web apps. We excluded web apps built with frameworks because we want to analyze the applications and not the frameworks.

For each app we collected as many versions as possible. Sometimes we could not get the whole lifecycle either because not all versions were available online or did not match the OOP criterion in the earlier versions.

The characterization of each web app included in our sample follows:

**PhpMyAdmin** (*https://www.phpmyadmin.net*) is an administration tool for MySQL and MariaDB. The initial release was on September 1998, but we only considered version 3 upwards, due to lack of OOP support and missing versions files.

**PhpPgAdmin** (*http://phppgadmin.sourceforge.net*) is an administration tool for PostgreSQL. Started as a fork of phpMyAdmin but has now a completely different code base.

**DokuWiki** (*https://www.dokuwiki.org*) is a wiki engine that works on plain text files and does not need a database.

**MediaWiki** (*https://www.mediawiki.org*) is a wiki engine developed for Wikipedia in 2002, and given the name "MediaWiki" in 2003. It was first released in January 2002.

**OpenCart** (*https://www.opencart.com*) is an online store management system, or e-commerce or shopping cart solution. Uses a MySQL database and html components. First release was on May 99.

**PrestaShop** (*https://www.prestashop.com*) is an e-commerce solution. Uses a MySQL database. Started in 2005 as *phpOpenStore*. Renamed in 2007 to *Prestashop*.

**PhpBB** (*https://www.phpbb.com/*) is an internet forum / bulletin board solution. Supports multiple database (*PostgreSQL, SQLite, MySQL, Oracle, Microsoft SQL Server*). Started in December 2000.

**Vanilla** (*https://open.vanillaforums.com/*) is a lightweight Internet forum package/solution. It was first released in July 2006.

---

[3] Note: PHP can be used with a pure procedural style; the object-oriented style became available from version 4 onwards.
[4] Domain database manager/administration tool, wikis, forums, shopping/stores

The complete list of applications is shown on Table 1. The LOC and Classes number are from the last version and were measured by the *PHPLOC*[5] tool.

*Table 1. Characterization of the target web apps*

| Name | Purpose | # of collected versions (collection period) | Last version | LOC (last version) | Classes (last version) |
|---|---|---|---|---|---|
| *PhpMyAdmin* | Database administration tool | 181 (09/2008 - 09/2019) | 4.9.1 | 301748 | 1174 |
| *DokuWiki* | Wiki solution | 40 (07/2005 - 01/2019) | 04-22b | 271514 | 402 |
| *OpenCart* | Shopping cart solution | 28 (04/2013 - 04/2019) | 3.0.3.2 | 206253 | 955 |
| *phpBB* | Forum/bulletin board solution | 50 (04/2012 - 01/2018) | 3.2.2 | 341159 | 1330 |
| *PhpPgAdmin* | Database administration tool | 29 (02/2002 - 09/2019) | 7.12.0 | 71210 | 54 |
| *MediaWiki* | Wiki solution | 145 (12/2003 - 10/2019) | 1.33.1 | 754941 | 2479 |
| *PrestaShop* | Shopping cart solution | 74 (06/2011 – 08/2019) | 1.7.6.1 | 516737 | 2597 |
| *Vanilla* | Forum/bulletin board solution | 75 (06/2010 – 10/2019) | 3.3 | 193435 | 533 |

## 2.2 Code smells sample

We used *PHPMD*[6], an open-source tool that detects scattered and localized CS in PHP. It only supports 3 scattered CS, so we chose the same number of localized ones, although much more of the latter are available. In the following paragraphs we characterize the selected CS. The first 3 types of are localized ones, i.e., they lie inside a class or file, and the last ones are scattered ones, because they spread across several classes.

**ExcessiveMethodLength** (aka "*Long method*"): a method has more lines than a predefined threshold, often associated with a standard screen size or programming style.

**ExcessiveClassLength** (aka "*God Class*"): a class may be trying to do too much.

**ExcessiveParameterList** (aka *"Long Parameter List"*): a method has a too long list of parameters.

**DepthOfInheritance**: a class is too deep in the inheritance tree; a class with many parents is an indicator for an unbalanced and wrong class hierarchy.

**CouplingBetweenObjects**: a class depends too much on other classes, which has a negative impact on several of its quality aspects such as stability, maintainability, and understandability.

**NumberOfChildren**: a class with an excessive number of children is an indicator for an unbalanced class hierarchy.

A brief characterization of all CS used is presented in Table 2. The thresholds used are the default ones used in PHPMD, with in turn came from PMD[7], and are generally accepted from the references in the literature[8] (Bieman & Kang, 1995; Lanza & Marinescu, 2007; Mccabe, 1976). These thresholds should be considered as baselines and could be optimized using an approach like the one proposed in (Herbold et al., 2011). The latter concludes that metrics' thresholds often depend on the properties of the project environment, so the best results are achieved with thresholds tailored to the specific environment. In our case, where we have 8 web apps, each developed by a different team, such optimization would lead to specific thresholds for each app, which would add confounding effects to the comparability between apps we want to carry out in this study.

---

[5] https://phpqa.io/projects/phploc.html
[6] https://phpmd.org/
[7] https://pmd.github.io/
[8] https://pmd.github.io/latest/pmd_java_metrics_index.html

*Table 2. Characterization of the target code smells*

| Code Smell | Characterization | Type | Threshold |
|---|---|---|---|
| ExcessiveMethodLength | The method does too much | Localized | 100 LOC |
| ExcessiveClassLength | The class does too much | Localized | 1000 LOC |
| ExcessiveParameterList | The method has too many parameters | Localized | 10 parameters |
| DepthOfInheritance | The class is too deep in the inheritance tree | Scattered | 10 levels |
| CouplingBetweenObjects | The class has too many dependencies | Scattered | 13 dependencies |
| NumberOfChildren | The class has too many descendants | Scattered | 15 descendants |

**2.3　Data collection and preparation workflow**

The workflow of our study (see Fig. 1) included a data collection and preparation phase before the data analysis phase and was fully automated by means of several tools. The following steps were performed in the data collection and preparation phase:

- Using a browser, the source code of all versions of the selected web apps was downloaded, mainly in ZIP format, from *GitHub, SourceForge,* or private repositories, except the alpha, beta, release candidates, and corrections for old versions, i.e., everything out of the main branch. When we had 2 branches in parallel, we considered only the stable branch; during this step, we created a database table with the application versions, later exported to a CVS file, containing the timestamps for each downloaded version.
- The downloaded ZIP files (one per each version of each app) were unzipped to flat files on the file system of our local computer (one folder per each version of each app).
- *PHPMD* extracted the location, dates, and other CS indicators from all versions and stored them in XML file format (one file per each version of each app). For the applications we excluded some directories that were not part of the applications (vendor libraries, images, etc.).
- The *CodeSmells2DB* PHP script read the previous XML files and, after some format manipulation, stored the corresponding information in a *MySql* database. The data at this point was stored by version/smell.
- *PHPMyAdmin* read the CVS file with the versions' timestamps created in the first step and merged that info with one collected in the previous step by the *CodeSmells2DB* script.
- The *CSLong2CSSurv* read the information stored in the *MySql* database and transformed it into a format suitable for survival analysis by statistics tools. That includes the date when each code smell was first detected and, if that was the case, when it disappeared, either due to refactoring or because the code where it was detected was (at least apparently) removed. The results of this transformation were stored back in the *MySql* database. Then, a data completion step was performed, where censoring and survival periods were calculated. Finally, results were exported to CVS format (one file per each app), in preparation for the data analysis phase.
- Finally, *PHPLOC* was used to extract several code metrics from the source code of each version of each app, storing them in one CVS file per app, thereby concluding this data preparation phase.

The aforementioned "censoring" activity encompassed transforming the collection of detected CS instances for each version of each web app, to a table with the "life" of each instance, including the date of its first appearance, removal date (if occurred) and a censoring value meaning the following:

- Censored=1 $\Rightarrow$ the smell disappeared, usually due to a refactoring event;
- Censored=0 $\Rightarrow$ the code smell is still present at the end of the observation period.

For replication purposes, the collected dataset is made available to the community in CVS format[9].

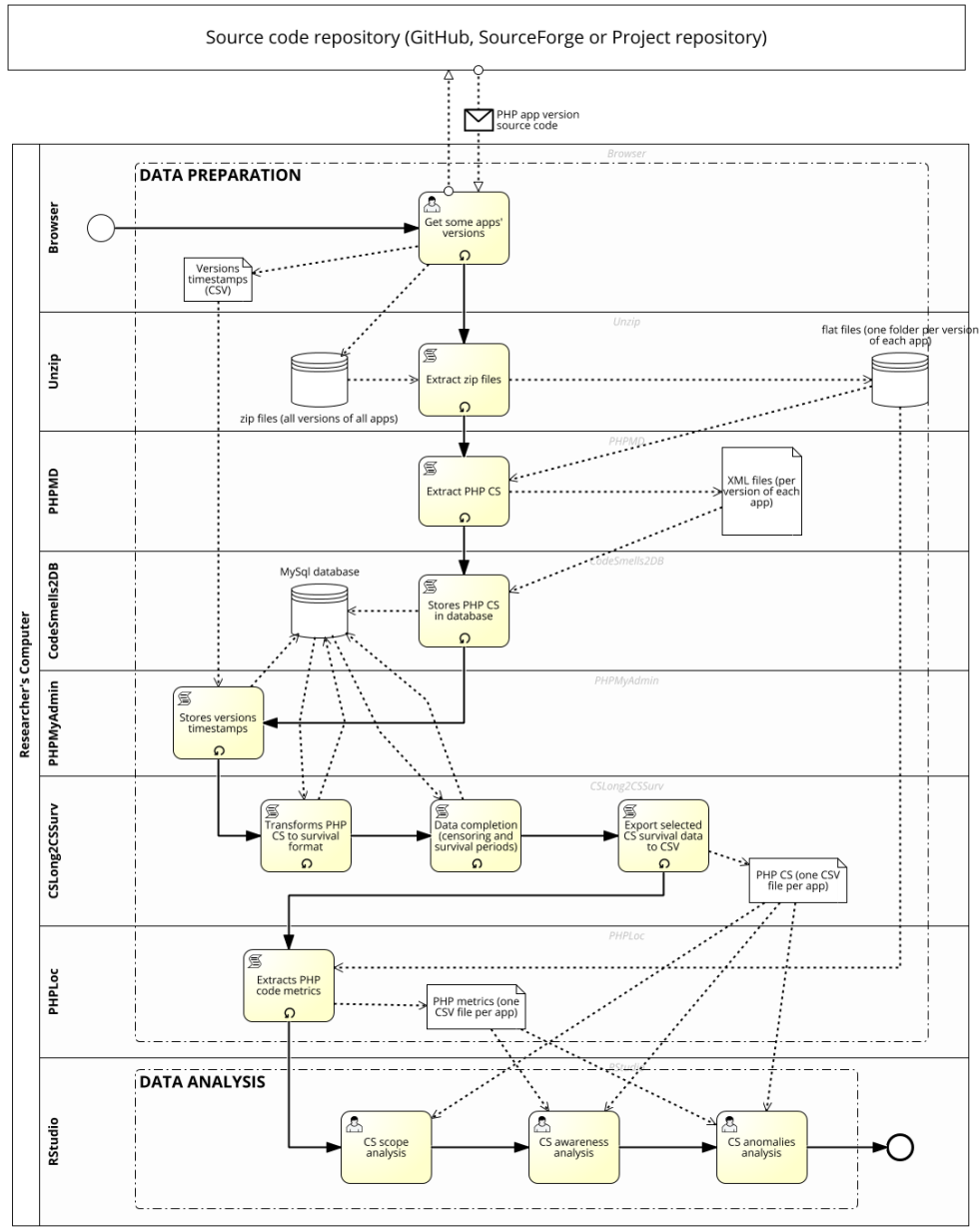

**Fig. 1.** *Workflow of the data preparation and analysis phases*

---

[9] https://github.com/americorio/articledata/

# 3 Survival analysis

Survival analysis encompasses a set of statistical approaches that investigate the time for an event to occur (Clark et al., 2003). The questions of interest can be the average survival time, and the probability of survival at a certain point in time. Also, we can calculate the hazard function, i.e., the probability of the event to occur.

To answer the first two research questions, we formulated and tested two hypotheses. We used the R statistics tool, with packages *survival*[10], *survminer*[11], and *dplyr*[12]. We performed two studies, one for each hypotheses, using the *log-rank test* (Daniel Schuette, n.d.) and the two different co-variables (type and scope). The *log-rank test* is used to compare survival curves of two groups, in our case two types of CS. It tests the null hypothesis that survival curves of two populations do not differ by computing a p-value.

## 3.1 Graphical view of the life of code smells

We start our survival analysis by plotting the life of the chosen CS. These plots are useful to visualize the life of each code smell.

### 3.1.1 Evolution of code smells by time in years

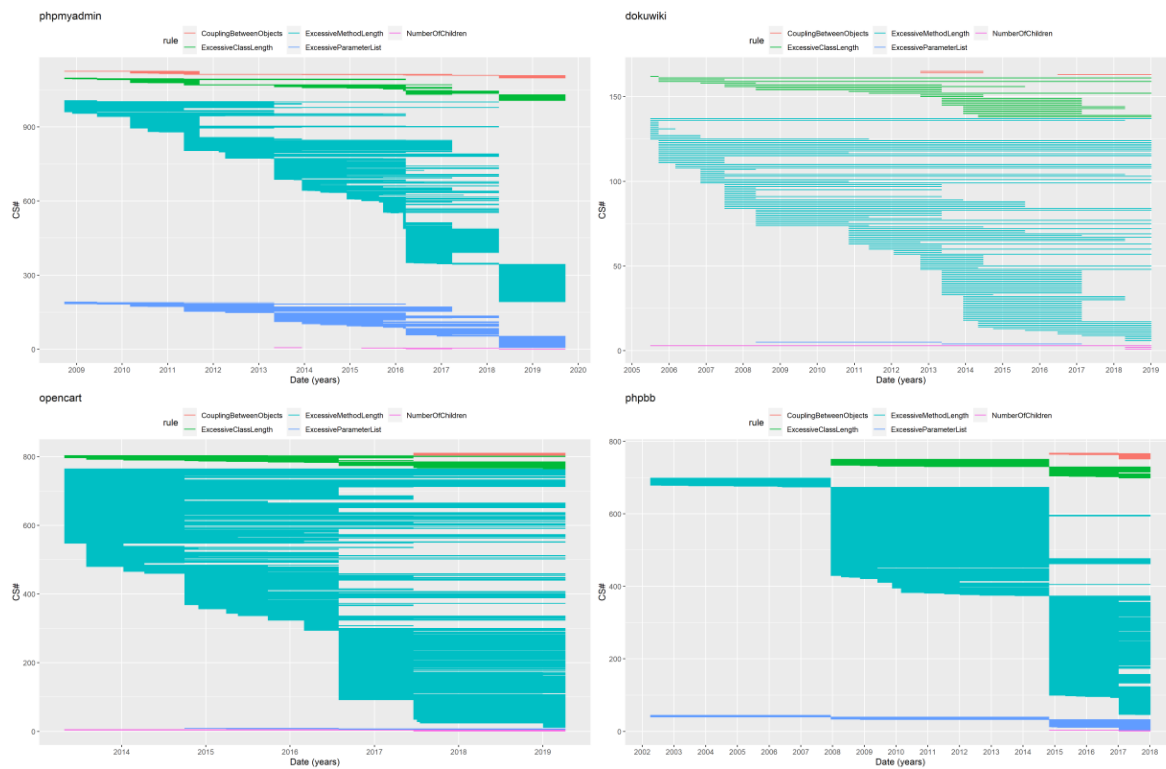

---

[10] https://cran.r-project.org/web/packages/survival/

[11] https://cran.r-project.org/web/packages/survminer/

[12] https://cran.r-project.org/web/packages/dplyr/

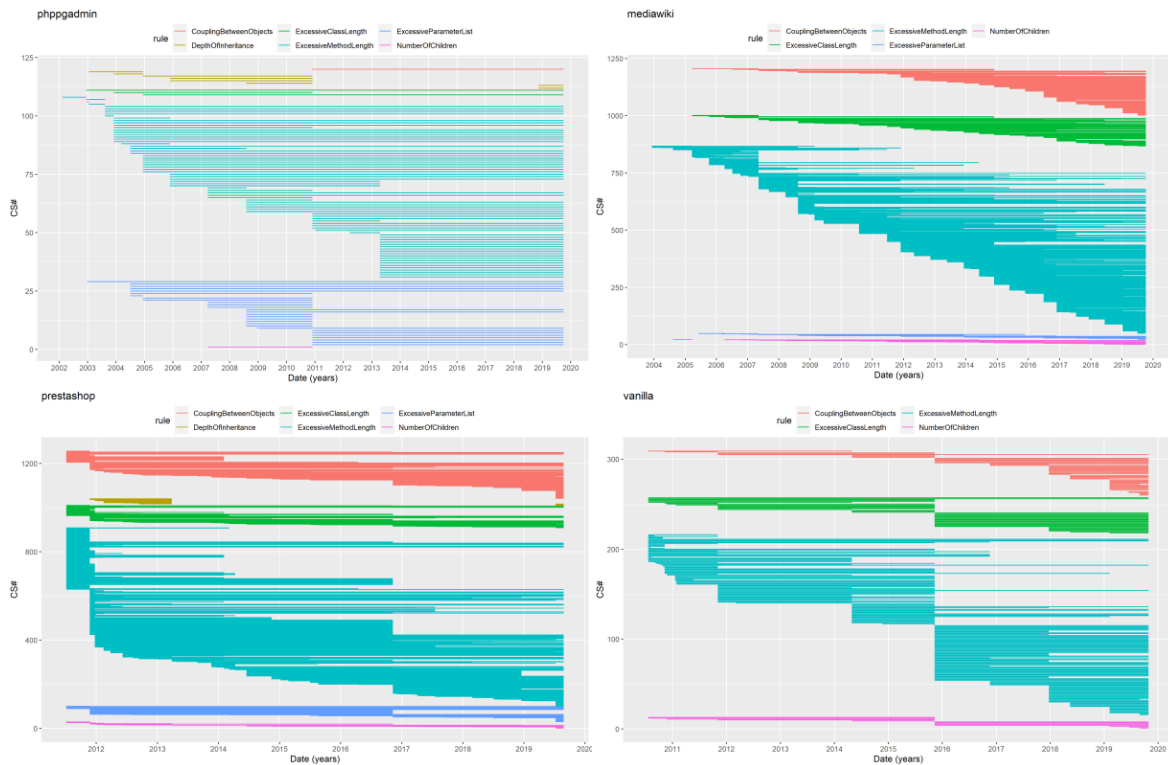

***Fig. 2.*** *Chronological evolution of the number of code smells, by application*

Fig. 2 represents the lifeline of each CS, from its introduction until its removal or the end of the collection period (what happens first). This allows us to identify when refactoring occurred, as we will discuss later. The "ExcessiveMethodLength" (aka "Long Method") smell is the more recurrent within the chosen set.

### 3.1.2 Evolution of code smells by version

Software delivery occurs at unequally spaced moments in time, that we usually call "versions". The history of versions is therefore an irregular time series. Fig. 3 represents the "version-span" of each CS, by application.

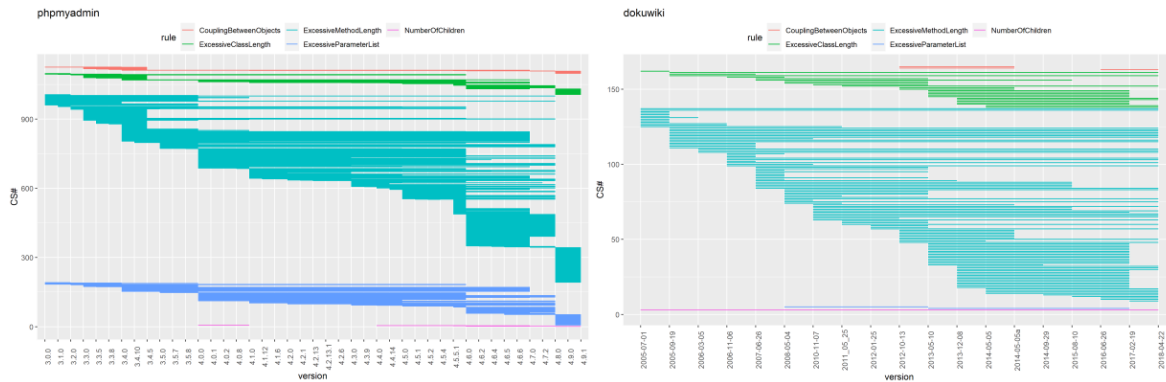

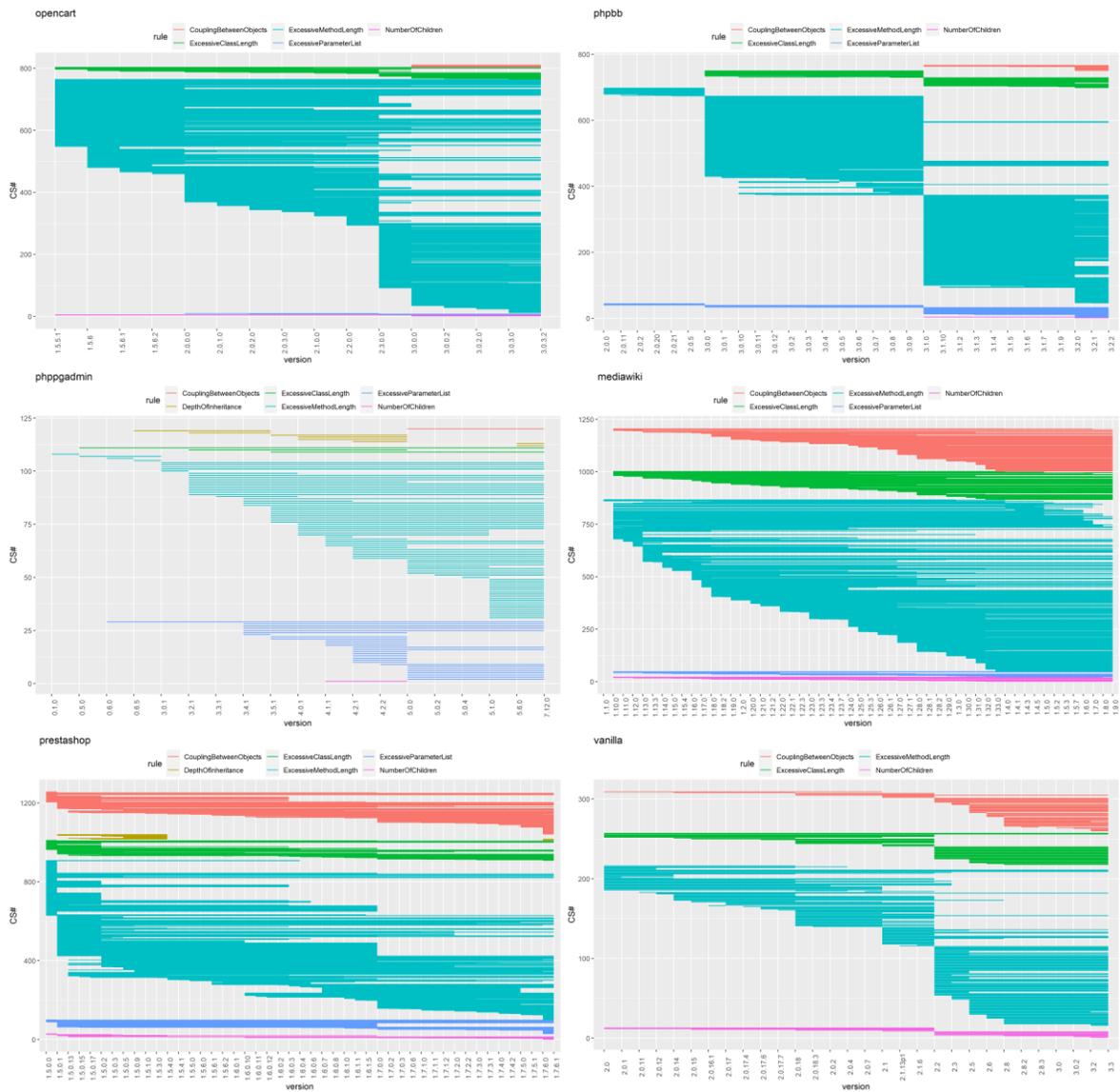

*Fig. 3. Evolution of code smells, by application, by version*

The previous visualization allows identifying the versions where a major refactoring occurred. For example, in *phpMyAdmin*, versions 4.6, 4.7 and 4.8 seem to have a lot of refactoring. We drilled down this behavior in version 4.8 because a lot of smells were introduced in this version. By code inspection we found that during refactoring some files changed their names, so the smell is considered new. In other words, a rename in a file or class causes the fake conclusion that some existing smells were removed while new ones were created. To block this fake effect, we should observe the evolution of the total number of occurrences for each CS, by version.

### 3.1.3 Evolution of number of code smells by version

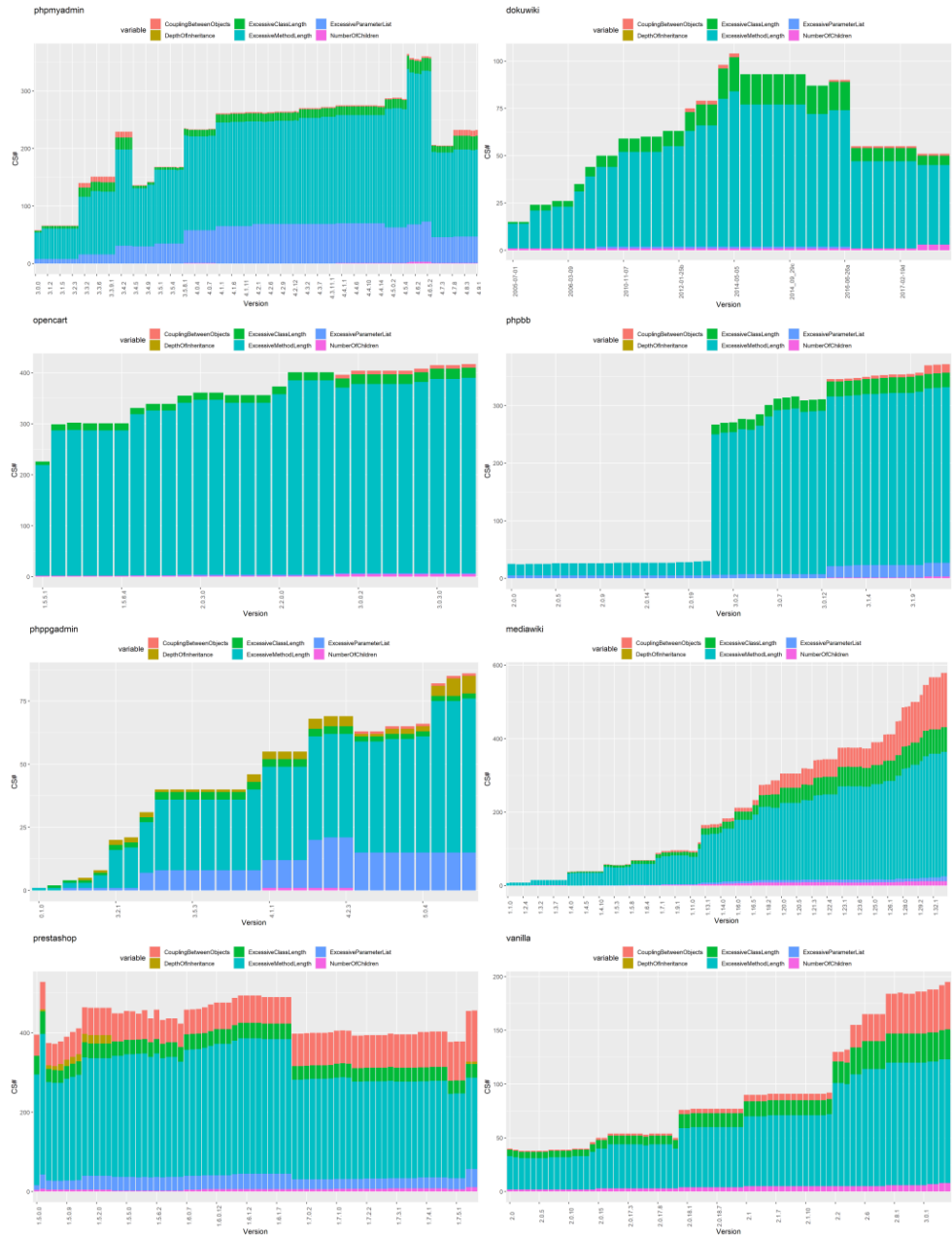

*Fig. 4. Evolution of total number of each code smell, by application, by version*

In Fig. 4, where the total amount of CS per application is shown, it is possible to identify the moments where intensive refactoring or large changes in the code base occurred. For instance, drilling down in the *phpMyAdmin* app, in addition to the removal of the CS, there was a replacement of some files that changed the name and directory. We observed these phenomena in the record of each individual CS. If we observe the Fig. 3, we don't understand if the smells are all new or came from renaming operations but observing the total number of code smells in Fig 4 we have a different perspective. We also pinpoint that for 2 of the applications, *vanilla* and *mediawiki*, the number of code smells increases steadily during the life of the application. We will discuss this later when we answer research question 4.

## 3.2 Survival curves for different types of code smells: "Localized" and "Scattered"

We now pose the following null hypothesis that derives from research question RQ1 (section 1.2):

> $H_01$: Survival does not depend on the code smells scope

To estimate the survival function and provide a probability that a given CS will subsist/exist past a time t, we will use the non-parametric Kaplan-Meier estimator (Kaplan & Meier, 1958). We fitted the Kaplan-Meier curves and performed the log-rank test to compute the p-value (statistical significance). A p-value less than 0.05 means that the survival curves are different between CS types with a 95% confidence level. Table 3 presents our findings, where the column names are self-explanatory.

Table 3. Code smells found, removed and survival in days (median and mean), by type.

| Web app | Type | Found | Removed | % removed | Median(d) | Mean(d) |
|---|---|---|---|---|---|---|
| phpmyadmin | Localized Smells | 1091 | 870 | 80% | 746 | 868 |
|  | Scattered Smells | 34 | 23 | 68% | 556 | 628 |
| dokuwiki | Localized Smells | 159 | 114 | 72% | 1381 | 2117 |
|  | Scattered Smells | 6 | 2 | 33% | 620 | 2779 |
| opencart | Localized Smells | 798 | 395 | 49% | 1189 | 1275 |
|  | Scattered Smells | 12 | 0 | 0% | NA | 2172 |
| phpbb | Localized Smells | 747 | 395 | 53% | 2512 | 2137 |
|  | Scattered Smells | 20 | 2 | 10% | NA | 1845 |
| phppgadmin | Localized Smells | 110 | 32 | 29% | NA | 3627 |
|  | Scattered Smells | 10 | 7 | 70% | 1589 | 1723 |
| mediawiki | Localized Smells | 979 | 577 | 59% | 1387 | 1954 |
|  | Scattered Smells | 226 | 70 | 31% | 2794 | 2712 |
| prestashop | Localized Smells | 981 | 690 | 70% | 1087 | 1271 |
|  | Scattered Smells | 275 | 134 | 49% | 943 | 1440 |
| vanilla | Localized Smells | 244 | 119 | 49% | 1469 | 1581 |
|  | Scattered Smells | 65 | 17 | 26% | 1469 | 1535 |

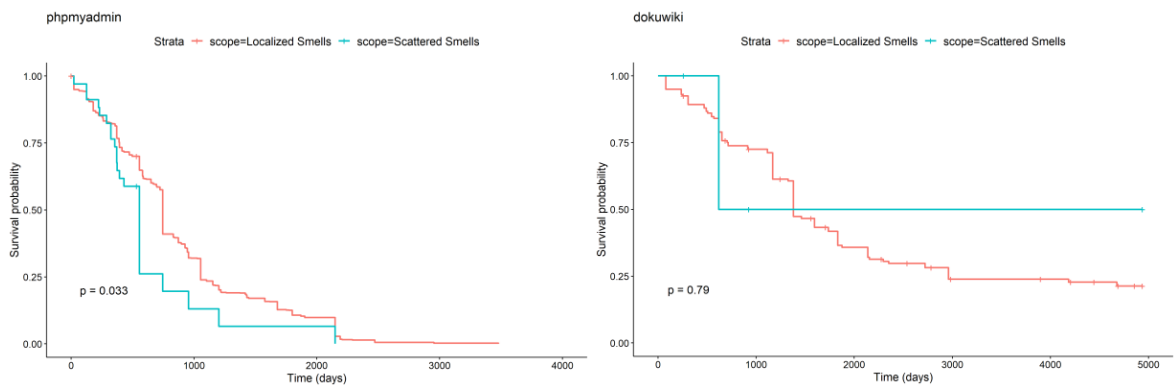

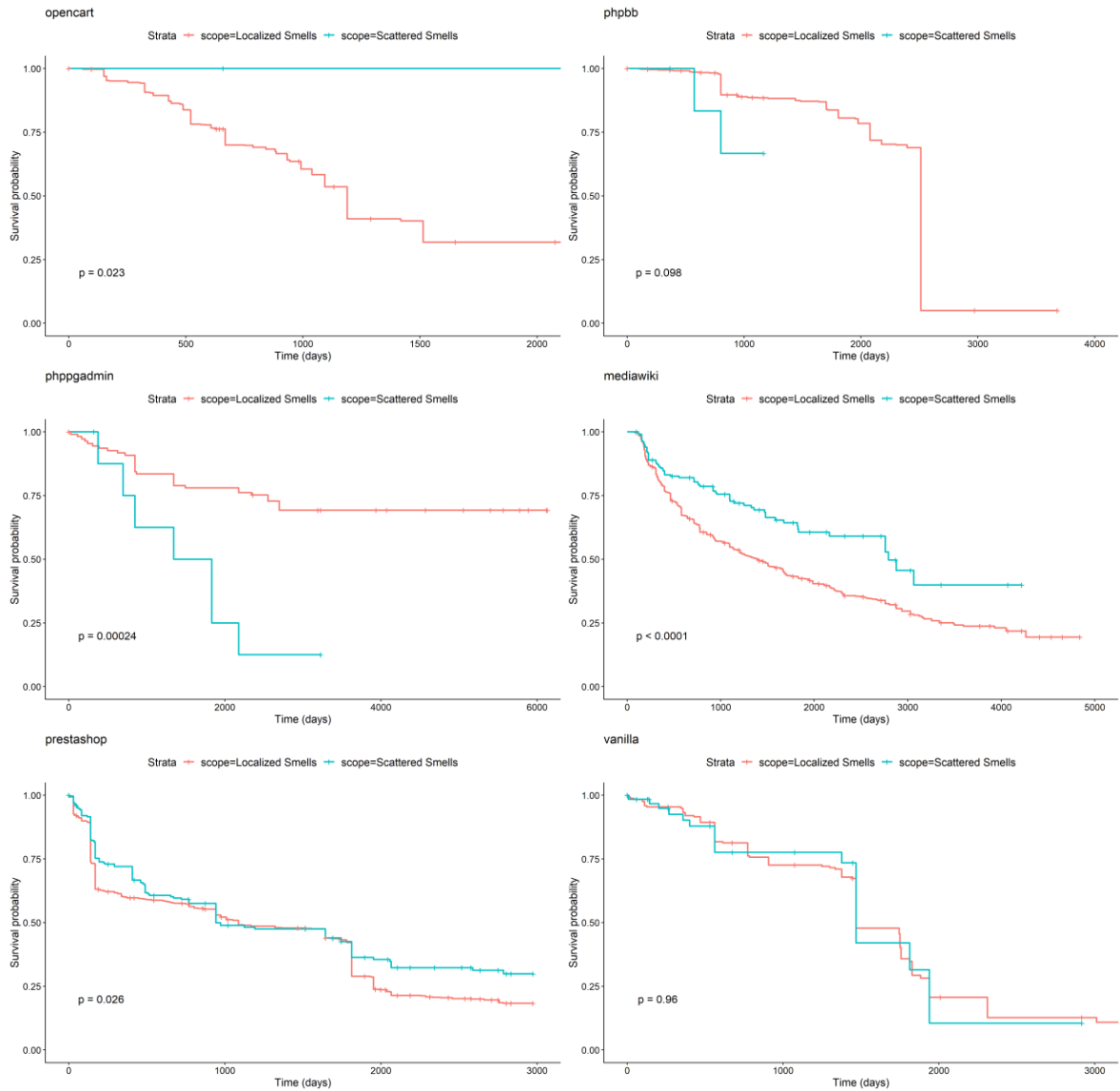

**Fig. 5.** Survival curves of localized and scattered code smells, by application.

Considering again a confidence interval of 95%, we can conclude from Table 4 that for *phpMyAdmin, opencart, mediawiki, phppgadmin* and *prestashop,* scattered and localized smells exhibit a different behavior regarding their survival curves. We also can see that for *phpbb* these curves are different considering a confidence interval of 90%. The differences for *dokuwiki* and *vanilla* are not statistically significant.

*Table 4.* p-value and Metrics for the last version of the application.

| Web App | Statistics | Github app metrics | | | Metrics | | | | | |
|---|---|---|---|---|---|---|---|---|---|---|
| | p | contributors | commits | releases | LOC | LLOC | Classes | Outside Classes and Functions | Percentage Outside Classes and Functions | LLOC per Contributor |
| phpmyadmin | **0,033** | 1032 | 116779 | 453 | 301748 | 66364 | 1174 | 9628 | 14.5% | 64,306 |
| dokuwiki | 0,79 | 523 | 10431 | 116 | 271514 | 75307 | 402 | 38996 | 51.8% | 143,990 |
| opencart | **0,023** | 274 | 9514 | 41 | 206253 | 75189 | 955 | 11790 | 15.7% | 274,412 |
| phpbb | 0,098 | 189 | 33336 | 155 | 341159 | 61575 | 1330 | 6922 | 11.2% | 325,794 |
| phppgadmin | **0,0002** | 28 | 2256 | 45 | 71210 | 36512 | 54 | 22295 | 61.1% | 1304,000 |
| mediawiki | **0,0001** | 514 | 93052 | 337 | 754941 | 157202 | 2479 | 7478 | 4.8% | 305,841 |
| prestashop | **0,026** | 581 | 57236 | 106 | 516737 | 102179 | 2597 | 19471 | 19.1% | 175,868 |
| vanilla | 0,96 | 114 | 28008 | 107 | 193435 | 48820 | 533 | 6602 | 13.5% | 428,246 |

On Table 4 we included some metrics taken from *GitHub* (the number of contributors, number of commits, and number of releases of each apps until the time analyzed) and also metrics measured with phpLOC (lines of code, logical lines of code, number of classes, code outside classes or functions, and it's percentage to the number of lines, and number of logical lines per contributor) for the last version of all applications, so we can analyze the 2 applications referred that are different form the others. For *dokuwiki*, only 2 scattered smells are removed (Table 3). Analyzing the outliers in the columns, we see that *Vanilla* and *phppgadmin* have high *LLOC per contributor.* Because of a small team and the size of application *phppgadmin* seems to have similar characteristics as the other applications, possibly because the refactoring would be easier for small applications. However, for *Vanilla* this can affect the results being studied, in the way that there is probably less refactoring, and we can observe that in the Fig 3. The reason why *opencart* curves are different, is because there is no removal of scattered smells during the observation period. Another factor that influences these results is the percentage of removal of scattered smells (Table 3). If this percentage is very small, it can lead to inconclusive results.

It is worth noticing that localized CS are much more frequent targets for change than scattered CS. This may be due to the lack of refactoring tools for scattered code smells PHP and the fact that manually removing scattered CS is much harder than for localized ones.

### 3.3 Survival curves for different time frames

In this section we now pose the following second null hypothesis that derives from research question RQ2 (see section 1.2):

> $H_0 2$: Survival of a given code smell does not change over time

We also used the *log-rank test*, and created a co-variate *timeframe*, with two values 1 and 2, 1 for the first half of the collection period, and 2 for the second half. For the first period, we truncated the variables of the study as if they were in a sub-study ending in this period. Therefore, in the end of the first period, we also filled the value of the *censored* column. The Table 5 contains the values found.

**Table 5.** Code smells found, removed and survival in days (median and mean) by timeframe.

| Web app | Timeframe | Found | Removed | % removed | Median(d) | Rmean(d) |
|---|---|---|---|---|---|---|
| *phpmyadmin* | 1 (< 2014-03-26) | 510 | 433 | 85% | 946 | 985 |
|  | 2 (>=2014-03-26) | 615 | 383 | 62% | 746 | 611 |
| *dokuwiki* | 1 (< 2012-04-03) | 94 | 55 | 59% | 2007 | 1599 |
|  | 2 (>= 2012-04-03) | 71 | 52 | 73% | 1381 | 1365 |
| *opencart* | 1 (< 2016-04-18) | 496 | 259 | 52% | 992 | 814 |
|  | 2 (>= 2016-04-18) | 314 | 35 | 11% | NA | 1007 |
| *phpbb* | 1 (< 2010-02-19) | 336 | 317 | 94% | 2512 | 2359 |
|  | 2 (>= 2010-02-19) | 431 | 80 | 19% | 1703 | 1464 |
| *phppgadmin* | 1 (< 2010-12-05) | 96 | 38 | 40% | NA | 2434 |
|  | 2 (>= 2010-12-05) | 24 | 1 | 4% | NA | 3084 |
| *mediawiki* | 1 (< 2011-11-07) | 512 | 401 | 78% | 779 | 1291 |
|  | 2 (>= 2011-11-07) | 693 | 222 | 32% | NA | 1987 |
| *prestashop* | 1 (< 2015-07-31) | 1008 | 565 | 56% | 943 | 817 |
|  | 2 (>= 2015-07-31) | 248 | 28 | 11% | NA | 1243 |
| *vanilla* | 1 (< 2015-03-10) | 130 | 82 | 63% | 1469 | 1152 |
|  | 2 (>= 2015-03-10) | 179 | 20 | 11% | NA | 1515 |

The survival curves for the 8 applications, and for the two timeframes, are represented in Fig. 6.

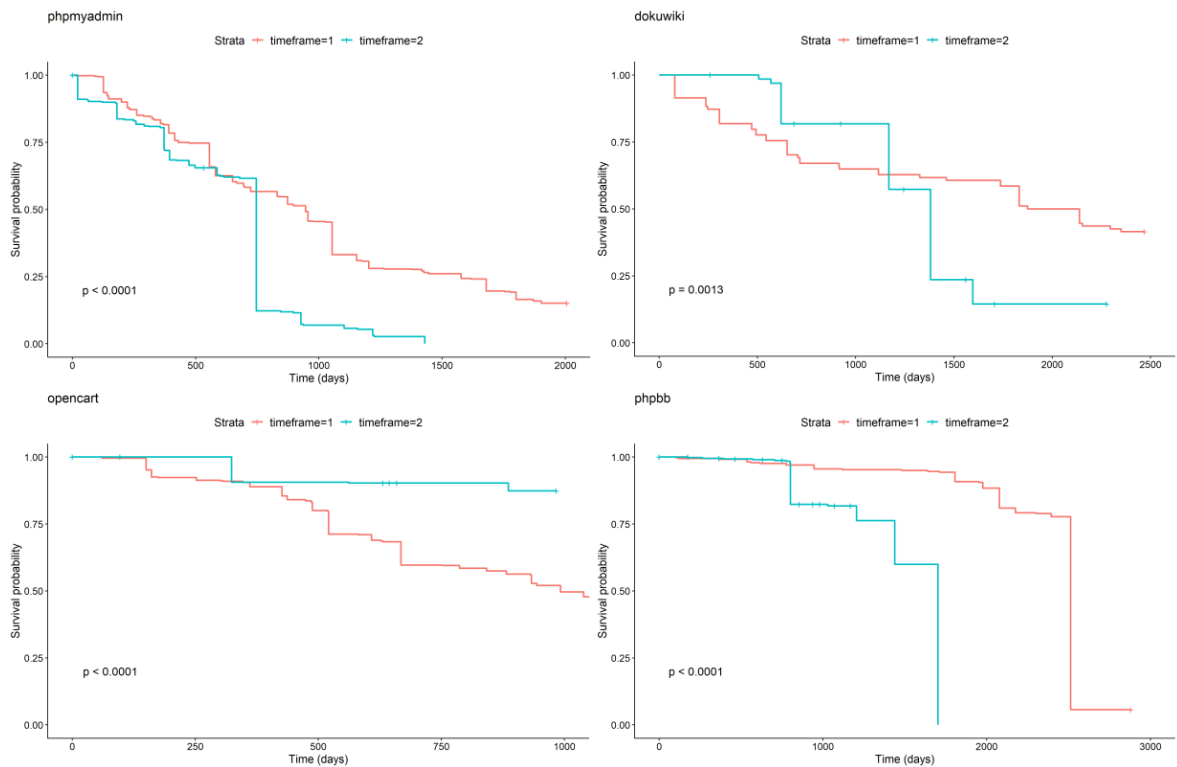

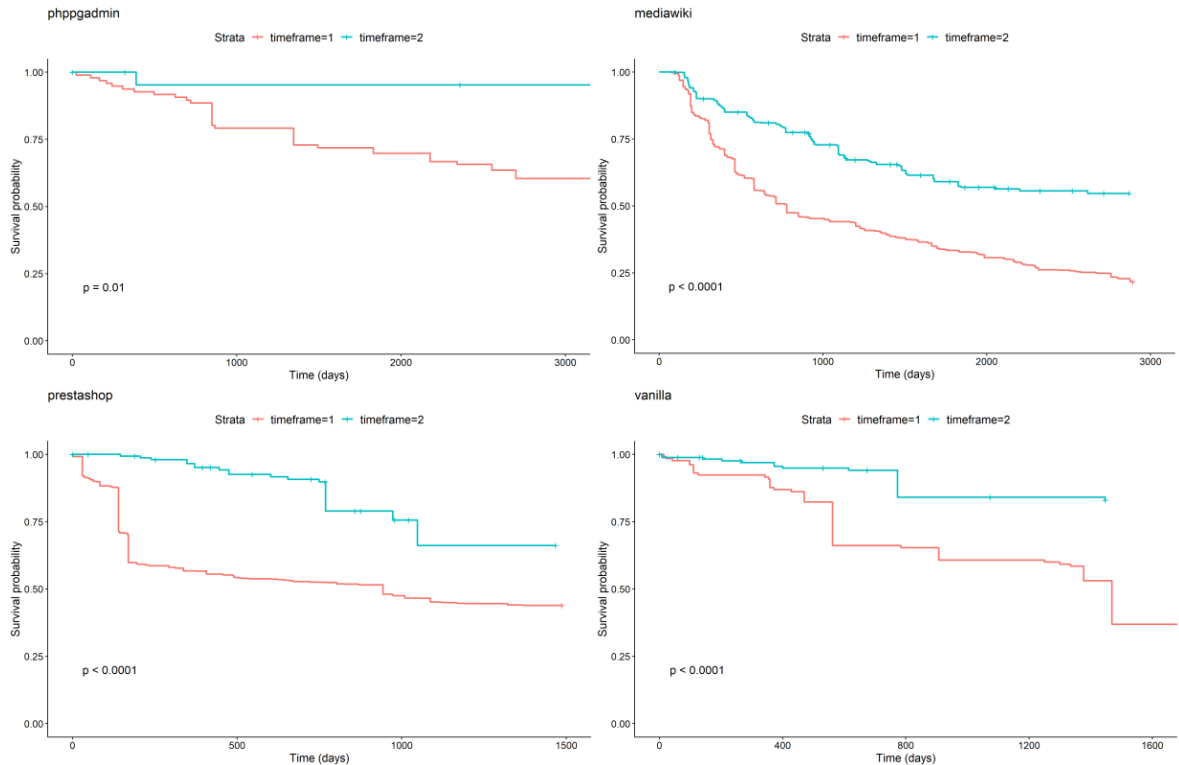

**Fig. 6.** *Code smells survival curves in two consecutive timeframes (1$^{st}$ and 2$^{nd}$ half)*

For all projects, the survival curves of the 1$^{st}$ timeframe differ significantly from the 2$^{nd}$ one. In *phpMyAdmin*, *dokuwiki* and *phpbb* the area under the CS survival curves is smaller in the 2$^{nd}$ timeframe, what seems to corroborate our expectation that, due to the increasing awareness on the potential harmfulness of CS, we would observe a reduction in the CS survival rate in the long term. However, for the others we cannot draw the same conclusion, i.e., the survival curves are different indeed, but the timeframe 1 has a greater area below the curve, so there are other factors to take in account.

On Table 6 we show the metrics in the end of timeframes 1 and 2, and the computed change rates (percentages of change).

**Table 6.** Metrics for the last version of timeframe 1 and last version of timeframe 2.

| | End Timeframe 1 | | | End Timeframe 2 | | | Δ (Change rate) | | |
|---|---|---|---|---|---|---|---|---|---|
| WebApp | LOC | LLOC | Classes | LOC | LLOC | Classes | Δ LOC | Δ LLOC | Δ Classes |
| phpmyadmin | 204496 | 46753 | 225 | 301748 | 66364 | 1174 | 0.48 | 0.42 | 4.22 |
| dokuwiki | 166036 | 54402 | 287 | 271514 | 75307 | 402 | 0.64 | 0.38 | 0.40 |
| opencart | 174047 | 70881 | 724 | 206253 | 75189 | 955 | 0.19 | 0.06 | 0.32 |
| phpbb | 198317 | 36821 | 212 | 341159 | 61575 | 1330 | 0.72 | 0.67 | 5.27 |
| phppgadmin | 96360 | 61465 | 43 | 71210 | 36512 | 54 | -0.26 | -0.41 | 0.26 |
| mediawiki | 955008 | 75174 | 1238 | 754941 | 157202 | 2479 | -0.21 | 1.09 | 1.00 |
| prestashop | 405968 | 94563 | 1023 | 516737 | 102179 | 2597 | 0.27 | 0.08 | 1.54 |
| vanilla | 147053 | 42103 | 374 | 193435 | 48820 | 533 | 0.32 | 0.16 | 0.43 |

If we take the number of classes as a measure of software complexity, we can say that all web apps match Lehman's *2<sup>nd</sup> Law of Software Evolution*, that refers to the increasing complexity of software systems: *"as an E-type system[13] evolves, its complexity increases unless work is done to maintain or reduce it"* (Lehman, 1979).

The factor that contributes the most for the lesser area in the first timeframe seems to be the removal percentage of CS (see table 4). For some applications, several CS exist since the beginning of the observation period and were not removed, so we will have a phenomena of CS accumulation.

As we can see in Fig. 2 and Fig. 3, there are clear refactoring traces in specific versions of *phpMyAdmin*, *dokuwiki* and *phpbb*, characterized by sudden drops in the number of CS. This behavior characterizes teams that avoid the technical debt syndrome (Cunningham, 1992).

In the remaining 5 apps, a monotonous increase in the number of CS is observed, which matches the *7<sup>th</sup> Law of Software Evolution*, that refers to the declining quality of software systems: *"the quality of an E-type system will appear to be declining unless it is rigorously maintained and adapted to operational environment changes"* (Lehman, 1996). By not being aware of the potential harm inflicted by CS, their development teams are already, or will become, technical debt hostages.

The popular adage that prevention is better than cure applies here entirely since the less costly way of getting rid of creeping CS is avoiding their introduction. This requires continuous awareness, and CS detection techniques and tools play a very important role here, especially when embedded in IDEs.

As for CS cure, a few refactoring tools exist, integrated in several IDEs for PHP (*PHPstorm*[14], *NetBeans*[15], *Zend Studio*[16] and *Eclipse PHP*[17]), as well as command-line tools such as *scisr*[18] and *rephactor*[19]. Although manual refactoring is costly, especially for scattered CS (where we are not aware of the existence of refactoring tools) since encompasses program understanding and may introduce bugs in code as any other software maintenance action (Kim et al., 2006), it is surely a good investment in the medium to long term (Leitch & Stroulia, 2003) that is worth allocating resources for.

### 3.4 Php code smells lifespan

We want to know the median of the survivability of the CS, i.e., the value in days when the probability of removal of the smells is 50%. For all the 6 smells studies this value is around 4 years. We can see that the survivability of localized and scattered smells is different ($p<0.05$ on the right graph) and we present the values in the table. The median for localized smells is a little less than 4 years and the scattered smells is almost 5 years.

---

[13] According to Martin Lehman, an *E-type* software system is a program written to perform a real-world activity; how it should behave is strongly linked to the environment in which it runs, and such a program needs to adapt to varying requirements and circumstances in that environment; all web apps considered in this study match this type definition.
[14] https://www.jetbrains.com/phpstorm/
[15] https://netbeans.org/
[16] https://www.zend.com/products/zend-studio
[17] https://www.eclipse.org/pdt/
[18] https://iangreenleaf.github.io/Scisr/
[19] http://rephactor.sourceforge.net/

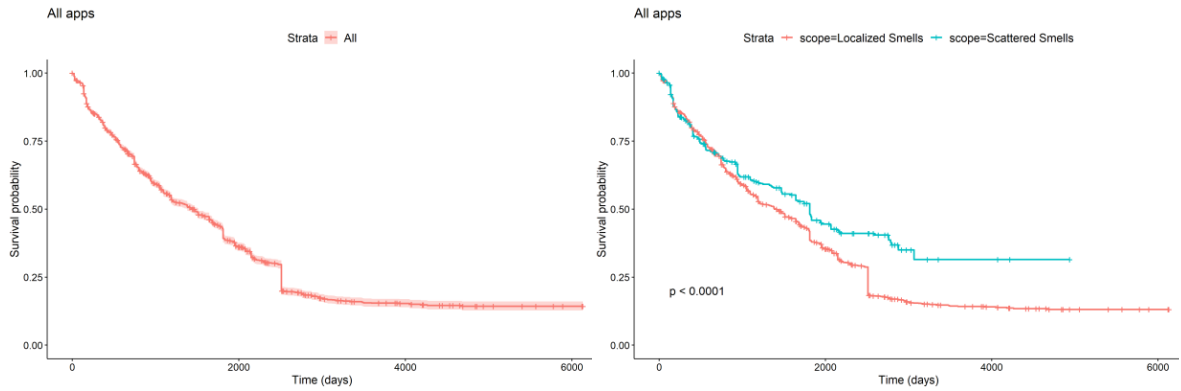

Figure 7 – Left: Survivability for all applications. Right: Survivability for all applications, by scope. P < 0.05 means the survivability of localized smells is different from scattered smells.

Table 7. Values of survivability of code smells, all applications and by scope (type)

| Smells | Num smells | Removed | % removed | median | rmean | se(rmean) |
|---|---|---|---|---|---|---|
| Total (all apps) | 5757 | 3447 | 60% | 1458 | 1979 | 34 |
| Localized | 5109 | 3192 | 62% | 1418 | 1924 | 34 |
| Scattered | 648 | 255 | 39% | 1812 | 2718 | 174 |

From Table 7 we can see also that in average 60% of smells are removed. However, if we take in account the scope, 62% of localized smells are removed against 39% of the scattered smells.

In Figure 8 we check if the survivability is the same for all the application using the 6 surrogate smells. We present it for each application, this time a graph with different curves from RQ1, due to having both types of smells grouped. We can see that survivability is not the same.

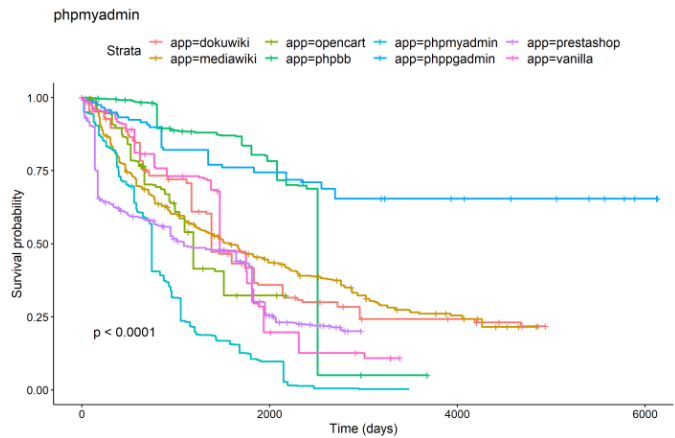

Figure 8 – Survivability of each app

In Table 8 we show the median for each application. We can guess that the applications with the median greater than the average median, do not remove the CS in a continuous mode (ex: *phpbb*, and in less degree, *mediawiki* and *vanilla*).

*Table 8 - Median of survival for each app*

| App | N | events | median |
|---|---|---|---|
| dokuwiki | 165 | 116 | 1381 |
| mediawiki | 1205 | 647 | 1583 |
| opencart | 810 | 395 | 1189 |
| phpbb | 767 | 397 | 2512 |
| phpMyAdmin | 1125 | 893 | 746 |
| phppgadmin | 120 | 39 | NA |
| prestashop | 1256 | 824 | 1087 |
| vanilla | 309 | 136 | 1469 |

We also show the boxplots for the lifespan of CS for each application. We can see that for almost all the applications the scattered smells have a greater life. However, for some applications, as we have seen before, the scattered smells do not get removed so often, and previous sentence does not hold true and depends on the awareness of CS.

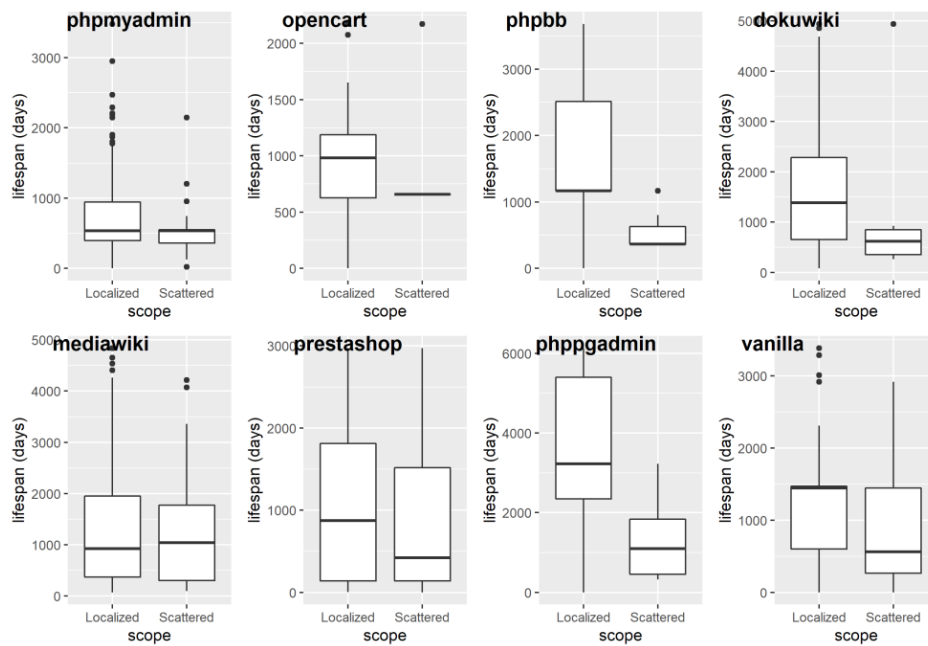

*Figure 9 - Boxplots of lifespan in days for the applications smells, localized, and scattered*

### 3.5 Threats to validity

There are four types of threats that can affect the validity of our experiments, we will see them in detail.

Threats to construct validity concern the statistical relation between the theory and the observation, in our case the measurements and treatment to the data. We detected the CS using PHPMD, where we detected 18 smells but only select 6 for the study as representatives / surrogates, to have 3 localized and 3 scattered, to have an equal number of smells on the two groups. For the second hypotheses we could have included all the smells detected.

The passage of data from timeseries to the format of survival format, with an initial date and a removal date is done by a program develop by us, that was tested for random smells and in the case that a lot of smells are removed in the same version. While we found no errors in the tests, it is worth mentioning.

Threats to internal validity concern external factors we did not consider that could affect the variables and the relations being investigated. We can say that PHPMD allows to change the thresholds of the of the CS detection, but we worked with the default values. These values can however be questioned for different applications.

Threats to conclusion validity concern the relation between the treatment and the outcome. In the survival studies we pose two hypotheses, checking for differences between two groups, and answer them with statistical support and significance. After the answers, we try to give measurements and causes for why one is higher than the other. In the anomaly detection part, we propose a simple way to pinpoint anomalies in evolution of density of the number of CS.

Threats to external validity concern the generalization of results. Here we try to choose PHP applications with support for classes, and that were not applications used to build other applications, like frameworks or libraries. Having 8 typical web applications makes the need to have more studies. For the best generalization it would be better if we used even more applications, and for the second study more smells.

## 4    Anomalies in code smells evolution

When we performed the exploratory study, we observed versions in which there was a refactoring on file names and location in folders, but the smells prevailed. They were considered new by our algorithm because they are in a different file/folder. So, is it possible to check those versions for anomalies in the number of CS? We start by showing the relative change of CS and size of the app.

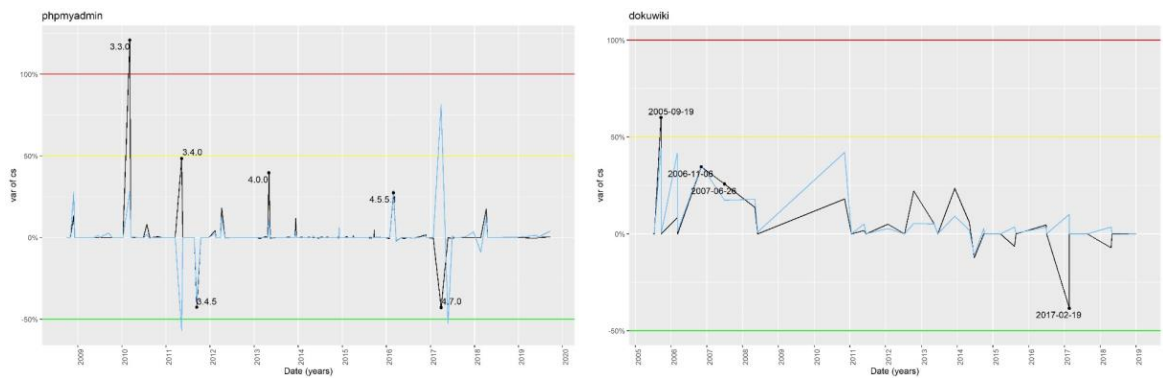

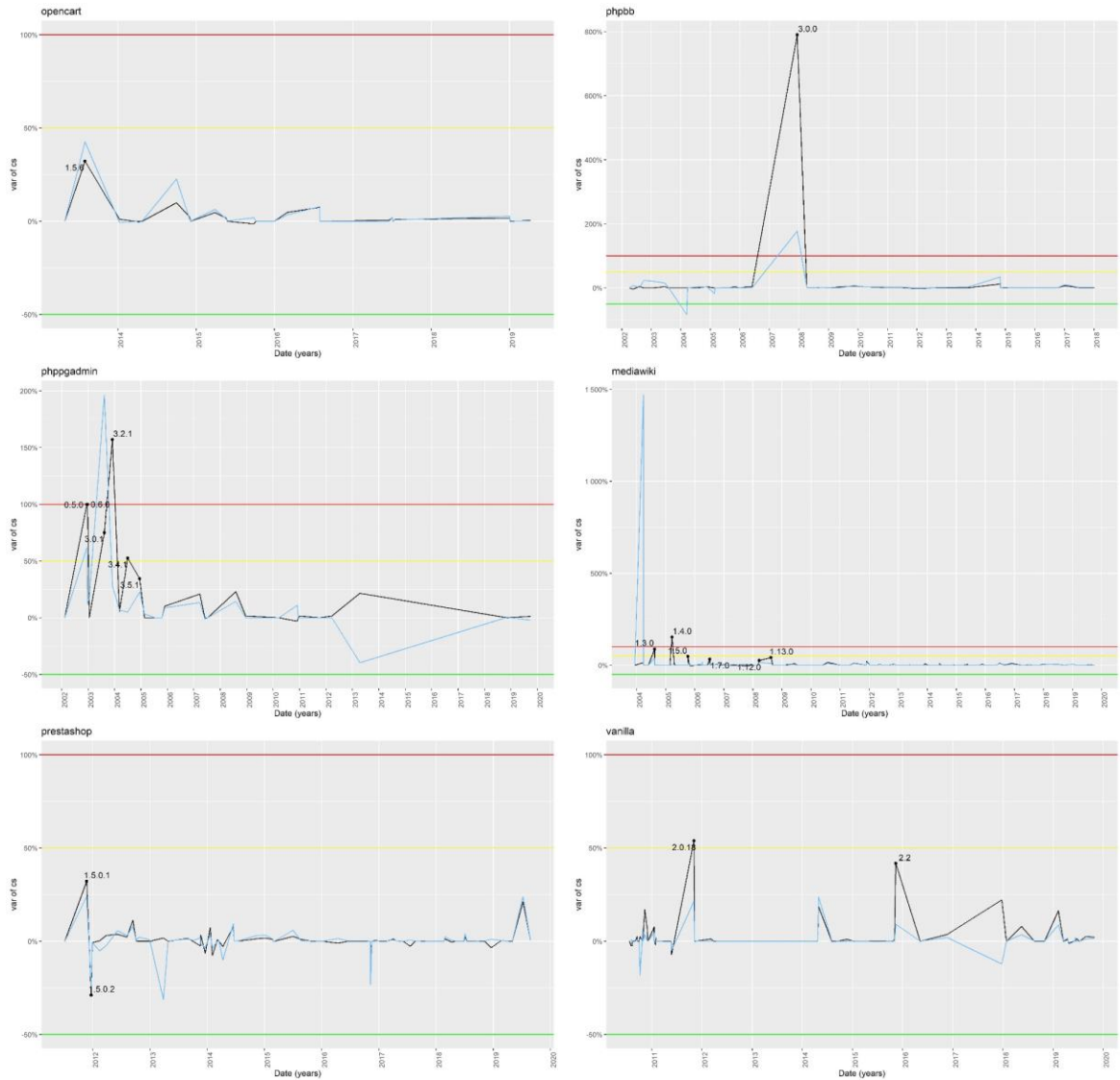

*Figure 10 Changes of cs (black) and KLLOC (blue)*

In this graphic we can see the relative change of CS from the previous version (black), and the relative change in KLLOC (thousands of logical lines of code) from the previous version. KLOC is a well know measure, although here we used the logical lines of code.

The relative change in number of CS is given by:

$$\Delta cs = \frac{cs_i - cs_{i-1}}{cs_{i-1}} = \frac{cs_i}{cs_{i-1}} - 1, \qquad (Eq.\ 1)$$

where $\Delta$ is the rate of change, $cs_i$ is the number of CS in the current version and $cs_{i-1}$ is the number of CS in the previous version. The same is calculated for the size, i.e. , the Logical lines of code.

The anomalies occur when there is a large increase in the number of CS and the size does not grow accordingly. It is also possible to get the version in which a lot of CS are removed by refactoring. But for a more direct way to spot anomalies, we can divide the #cs by the size. For the size measure we can use the Lines of Code, or Number of Classes (Henderson-Sellers, 1995).

We represented these measures of size in a graph (not shown here because of size constraints) and the (logical) lines of code is more appropriate to represent the size, because if we use the number of classes, we could be misrepresenting the size for the programs that have big classes (a CS). Therefore, we use CS density or ρcs=number of CS/LLOC.

We can now calculate the rate of change of the CS density, which we calculate in the same way as referred before for the CS number:

$$\Delta\rho cs = \frac{\rho cs_i - \rho cs_{i-1}}{\rho cs_{i-1}} = \frac{\rho cs_i}{\rho cs_{i-1}} - 1, \qquad \text{(Eq. 2)}$$

where $\Delta\rho cs$ is the rate of change of density of CS, $\rho cs_i$ is the density of CS in the current version and $\rho cs_{i-1}$ is the density of CS in the previous version.

We present the graph of this evolution of density, and this makes it easy to pinpoint the anomalies or outliers, that for easier visualization we show with a label.

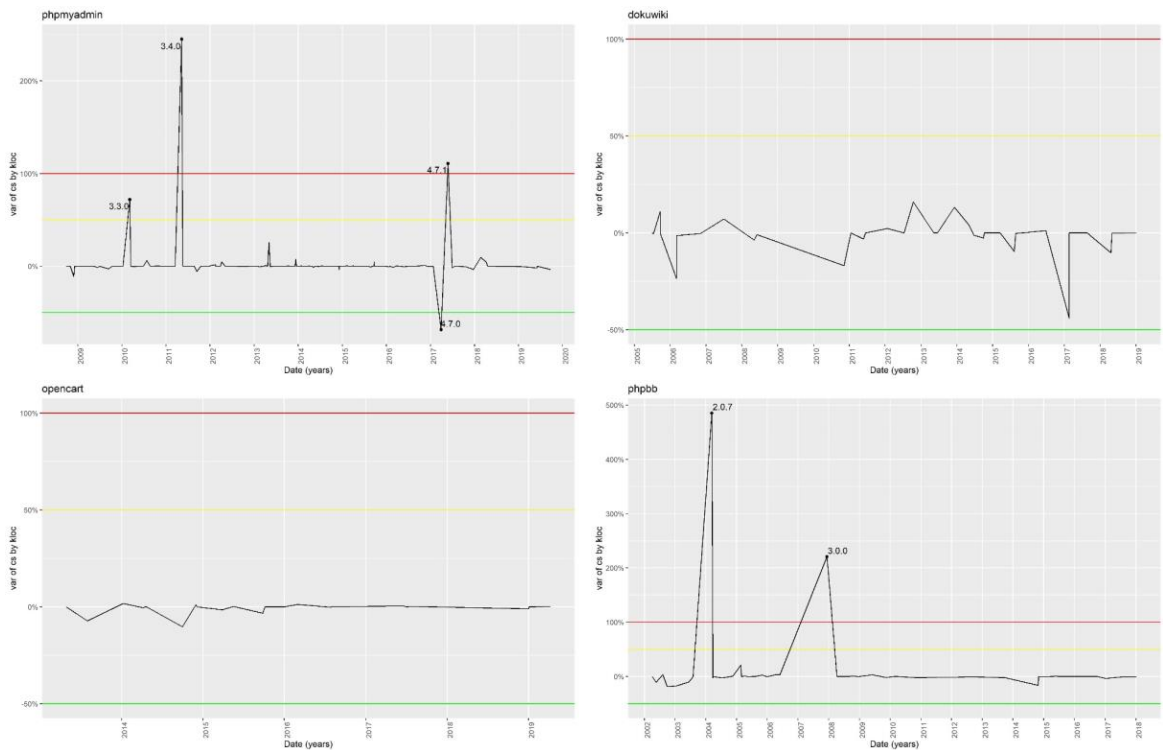

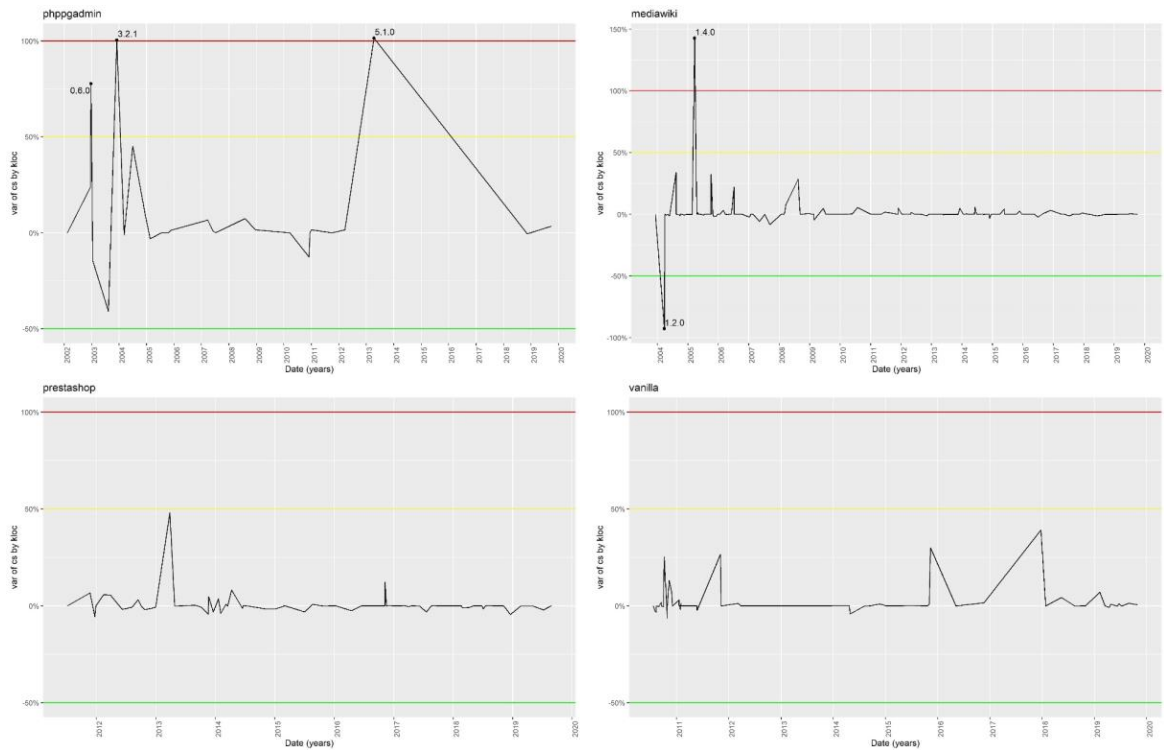
*Figure 11 Anomaly detection in number of code smells: changes of CS per changes of LLOC*

In the graphs per application, we use lines representing thresholds, signaling the increase of 50% and 100% and the reduction of 50% in the rate of change in the density of CS.

We can see that in 4 applications we find those anomalies. For example, phpMyAdmin has 3 of the anomalies, where the CS rise without the correspondent rise in the size of the program. This can indicate that the code has some problems in those versions. We also spot a decrease anomaly, in version 4.7, which can indicate that in this version there was some refactoring that reduce the number of CS. We can confirm this looking at figure 4.

This somewhat simple method to detect the anomalies in the CS by size can be put in a continuous development strategy, in which you make various tests before the release goes out, by using the percentage and having a limit to stop and correct if the number is above that. This can be all numeric or with a dashboard, with the suggested thresholds.

When we started this study of anomalies, when we saw the representation of the change of CS, we tried to apply SPC (statistical process control) with 2 or 3 standard deviations as limits, but we could not get limits due to the nature of the evolution (for long periods the value was the same, then sudden increases).

## 5   Related work

An extensive literature in software evolution has been published in the last decades. Since our current work is on evolution of CS in web apps, we will only review those papers that cover at least one of those topics. In other words, we will consider longitudinal studies on software evolution that treat either web apps or CS. We also present some related work in CS in web apps, that are not longitudinal studies. In another article with complementary work, we deal with detection of CS, and there we present the related work concerning their detection. We also show some work related to our questions specifically.

## 5.1 Longitudinal studies with web apps, but not on code smells

In (Kyriakakis & Chatzigeorgiou, 2014), the authors study five web apps in PHP, the aspects of their history, unused code, the removal of functions, the use of libraries, the stability of interfaces, migration to object-orientation and the evolution of complexity. They found these systems undergo systematic maintenance.

In (Amanatidis & Chatzigeorgiou, 2016), the authors analyze thirty PHP projects for their metrics, to examine if the Lehman's laws of software evolution could be observed in an web app, and found that not all of them hold.

## 5.2 Longitudinal studies on code smells, but not with web apps

### 5.2.1 Including survival techniques

A evolution study of CS is conducted by Chatzigeorgiou and Manakos (Chatzigeorgiou & Manakos, 2010), where they study 3 CS throughout a window of successive versions of 2 Java opensource systems. Later, they extend this work (Chatzigeorgiou & Manakos, 2014) to use 4 smells and survival analysis with survival curves. Their conclusions: in most cases, the CS persist up to the latest examined version thus accumulating. Survival analysis shows that smells "live" for many versions, being a permanent problem once introduced. A significant percentage of the CS was introduced in the creation of class/method. Very few CS are removed from the projects, and their removal was not from refactoring activities but a side effect of adaptive maintenance.

Tufano et Al. (Michele Tufano et al., 2015) conducted a large empirical study over the change history of 200 open source projects. They developed a strategy to identify smell-introducing commits, mining over half a million of commits, and manual analysis and classification of over 10K of them. Later they extended the paper (Michele Tufano et al., 2017) to include survival analysis. The findings show that most of the smell instances are introduced when an artifact is created and not because of its evolution, 80 percent of smells survive in the system, and among the 20 percent of removed instances, only 9 percent are removed as a direct consequence of refactoring operations.

In (M Tufano et al., 2016), after a survey to developers, the authors analyze when test smells occur in Java source code, what their survivability is, and whether their presence is associated with the presence of design problems in production code (CS). They found, among other conclusions, relationships between test and CS. They extracted data from a Git repository from 3 ecosystems making a total of 152 projects. They employ, among other techniques, survival analysis to study the lifespan of test smells.

In (Habchi et al., 2019), the authors present an empirical study that investigates the survival of (Java) Android CS, covering 8 types of Android CS, 324 Android apps, 255k commits, and the history of 180k code smell instances. Conclusions: a) CS can remain in the codebase for years before being removed; b) in terms of commits it takes 34 effective commits to remove 75% of them; c) Android

CS disappear faster in bigger projects with higher releasing trends; d) CS that are detected and prioritized by linters tend to disappear before other CS.

### 5.2.2 Not including survival techniques

Olbrich et al. study (S. Olbrich et al., 2009) focuses on the evolution of CS within a system and their impact on the change behavior (change frequency and size). They investigate two CS, analyzing the historical data over several years of development of 2 large open-source systems. They compare increase and decreases of classes infected with CS and total classes, in windows of 50 commits in subversion. They also study the relation between the change frequency of infected with CS classes and not infected. The results show different phases in the evolution of CS and that code smell infected components exhibit a higher change frequency. The information is useful for the identification of risk areas within a software system that need refactoring to assure a future positive evolution.

Tackling the claim that classes that are involved in certain CS are liable to be changed more frequently and have more defects than other classes in the code, in (S. M. Olbrich et al., 2010) they if this is true for God Classes and Brain Classes, with and without normalizing the effects with respect to the class size. they analyzed historical data from 7 to 10 years of the development of 3 systems. They used historical data from subversion and Bugzilla. Without normalization, the results show that God and Brain Classes were changed more frequently and contained more defects than other kinds of class. However, when they normalized the measured effects with respect to size, then God and Brain Classes were less subject to change and had fewer defects than other classes. They conclude that the presence of God and Brain Classes is not necessarily harmful and such classes may be an efficient way of organizing code.

Peters et Al. (Peters & Zaidman, 2012) study the lifespan of CS and refactoring behavior of developers in a software system by mining the software repository in 7 open source systems. They mine svn repositories and they built a semi-automatic tool to analyze the evolution of CS using eclipse detection plugins. Results indicate CS lifespan close to 50% of the lifespan of the systems; engineers are aware of CS, but are not very concerned with their impact, given the low refactoring activity. They also study the youngest 20% of smells and the latest 20% and they find that smells in the beginning of the life of the system are prone to be corrected quicker.

Rani et al. (Rani & Chhabra, 2017) perform an empirical study on distribution of different CS over different versions of projects, aiming to build refactoring strategies by testing which smell is more effective and at what time, during evolution of software. They study 4 versions of 3 software's. They compare the contribution of CS detected and refactored by Jdeodorant, with the overall status of smells with PMD. Study shows that a) Latest version of software has more design issues than that of oldest ones; b) "God" smell has more contribution for the overall status of CS, and "Type Checking" less. They also note that the first version of the software is cleaner.

Diglas et al. (Digkas et al., 2017) studied 66 Java open-source software projects on the evolution of technical debt over 5 years at the temporal granularity level of weekly snapshots. They calculate the trends of the technical debt time series and investigate the components of this technical debt. Their findings: a) technical debt together with source code metrics increase for most of the systems; b) technical debt normalized to the size of the system actually decreases over time in the majority of the systems c) that some of the most frequent and time-consuming types of technical debt are related to improper exception handling and code duplication.

Digkas et al. (Digkas et al., 2020) investigate the reasons to the introduction of technical debt, specifically: (a) the temporality of code technical debt introduction in new methods, i.e., whether the introduction of technical debt is stable across the lifespan of the project, or if its evolution presents spikes; and (b) the relation of technical debt introduction and the development team's workload in a given period. To answer these questions, we perform a case study on twenty-seven Apache projects, and inspect the number of Technical Debt Items introduced in 6-month sliding temporal windows. The results of the study suggest that: (a) overall, the number of Technical Debt Items introduced through new code is a stable metric, although it presents some spikes; and (b) the number of commits performed is not strongly correlated to the number of introduced Technical Debt Items.

### 5.3 Nonlongitudinal studies with code smells in web apps

In (Saboury et al., 2017), the authors address the faults in the releases of five JavaScript projects (1 framework, 2 libraries and 2 command line programs). They detect 12 JavaScript CS, and perform survival analysis, comparing the time until a fault occurrence, in files containing CS and files without CS. Results: (1) on average, files without CS have hazard rates 65% lower than files with CS. (2) Among the studied smells, two CS have the highest hazard rates. They also conduct a survey with JavaScript developers, and the assessment is in line with the findings of our quantitative analysis.

As an extension to the previous paper, in (Johannes et al., 2019) the authors perform a large-scale study of 12 JavaScript CS in 15 applications (libraries an frameworks, and 1 app), to understand how they impact the fault-proneness of applications, and how they are evolved by the developers of the applications. They perform survival analysis, comparing the time until a fault occurrence, in files containing CS and files without CS. They also examine the introduction and removal of the CS in the applications using survival models. The analysis is conducted at the granularity of the line of code. Results: (1) on average, files without CS have hazard rates at least 33% lower than files with CS. (2) Among the studied smells, 3 smells have the highest fault hazard rates. (3) CS, and particularly "Variable Re-assign," are often introduced in the application when the files containing them are created. Moreover, they tend to remain in the applications for a long period of time; "Variable Re-assign" is also the most prevalent code smell.

Amanatidis et al. (Amanatidis et al., 2017) investigate the relations between Technical Debt (that includes CS) and the maintenance/changeness of the files in PHP applications. They performed a case study on 10 open-source PHP projects, to assess the relation between the amount of TD and two aspects of interest: (a) corrective maintenance (i.e., bug fixing) frequency, which translates to interest probability and (b) corrective maintenance effort which is related to interest amount. They find that on average, the number of times that a high TD file is modified is 1.9 times larger than the number of times a low TD file is changed. In terms of the extent of change (number of lines), the corresponding ratio is 2.4 to 1. Conclusions: modules with a higher level of incurred TD, are more defect-prone and consequently require more (corrective) maintenance effort.

In (Bessghaier et al., 2020), the authors also conduct an empirical study to investigate PHP CS, a form of technical debt. They study 5 open-source web apps built with PHP as the server language, studying the diffuseness and their relationship with the change proneness of affected code. The key findings: 1) complex and large classes and methods are frequently committed in PHP files, 2) smelly files are more prone to change than non-smelly files, and 3) Too Many Methods and High Coupling are the most associated smells with files change-proneness.

The last 2 studies in PHP agree with studies in Java (Palomba et al., 2018) that report similar findings.

In (Aniche et al., 2018) and (Aniche et al., 2017) the authors present a catalog of six smells tailored to MVC Spring applications, defined by surveying/interviewing developers. They assess the relationship between the smells and the code change- and defect-proneness, investigate when these smells are introduced and how long they survive, and survey 21 Spring developers to verify their perception of the defined smells. Conclusions: the defined Web MVC smells (i) have more chances of being subject to changes and defects, (ii) are mostly introduced when the affected file is committed for the first time in the repository and survive for long time in the system, (iii) are perceived by developers as severe problems, and (iv) they try to generalize to other languages/frameworks interviewing experts.

In the work (Correia & Adachi, 2019), they present a catalog of 5 MVC smells tailored to Django web applications inspired by the previous work, and a tool to assist their automatic detection. They use different violations from the previous article that are not suitable for Django, especially because non-existent concepts in this framework (repository). They conducted an empirical study within an industry *Python* information system. They show how most recurrent violations evolve along software evolution, and the opinions and experiences of software architects regarding these violations.

They are more articles that deal with detection; however, we will treat them in a separate article, that investigates detection of CS.

### 5.4    Studies comparing types of smells

In this point we show studies comparing types of smells, namely measured at different levels.

In (Arcelli Fontana et al., 2019), the authors try to understand if architectural smells are independent from CS or can be derived from a code smell or from one category of them. The method used was analyzing correlations among 19 CS, 6 categories of CS, and 4 architectural smells. After finding no correlation, they conclude that they are independent of each other.

The paper (Sharma et al., 2020) aims to study architecture smells characteristics, investigate correlation, collocation, and causation relationships between architecture and design smells. They mined 3073 C# repositories and used 19 smells. Results: i) smell density does not depend on repository size; ii) architecture smells are highly correlated with design smells; iii) most of the design and architecture smell pairs do not exhibit collocation; iv) causality analysis reveals that design smells cause architecture smells (with statistical means).

The studies for Java and C# seem to have different results. This should require further investigation.

### 5.5    Studies with anomalies

We did not find a lot of relevant studies concerning anomalies in the evolution of CS. However, in a recent study from Diglas et al. (Digkas et al., 2020) already referred, they study the fluctuation in the evolution of technical dept (which includes CS), and they divide the applications in stable and sensitive (if they have spikes). To perform this classification, they use the SMF metric, i.e, the "software metrics fluctuation", which is defined as the average deviation from successive versions pairs.

Although we did not now this measure at the time of investigation, before settle in the way to detect anomalies in the CS, we explored a method with the standard deviation, and using control charts, and some more methods. However, the applications studied have several versions with no increase at all in the number of CS and LLOC, and some versions with a sudden increase, that, for those applications, we concluded that the method we presented is best suited to discover anomalies, also having the advantage of being very simple to implement in a quality pipe, because you do not need to know all the history of the project (see answer to RQ4).

### 5.6 Related works discussion

From related works presented, we can infer that the main techniques used by the authors to extract the code for further analysis are divided in: i) getting full releases from software repositories or sites; ii) mining Git (or other CVS, like subversion) repositories for the changes/commits. The second option can be automated and find connection with the smells and faults in the same file. However, this will work for the localized smells, the smells that are on one file. To compute scattered CS, i.e., smells that are present across various files, we had to deal with the full code of each version. For smells based only on metrics for the one file, the direct Git extraction is simpler.

Regarding studies with web applications, some of the studies use libraries, frameworks, and applications that generate other applications, in the application sample. This will bias the results if third-party code will also be considered in CS detection.

Another issue that we found is that some PHP applications used in the studies (e.g., *Wordpress*) are not fully object-oriented, which we do not deem adequate if the CS used were defined for that paradigm, such as the ones proposed in (Fowler, 1999).

About the relation of CS with changes of the files, there are some conclusion (S. Olbrich et al., 2009), (Faragó et al., 2014) about the impact in the quality that makes it a divided topic.

Although we have performed some forward and backward snowball, by no means we claim that our literature review was neither systematic, nor exhaustive. However, we are currently undertaking such an SLR (to be published soon) and our preliminary conclusions confirm that longitudinal studies on CS in web apps is a scarcely found topic in existing literature.

## 6  Conclusions and future work

The literature is still scarce in what concerns PHP CS evolution studies, the main topic of this paper. PHP, a language that fully supports object oriented paradigm (among others), is by far the most used on the server-side for web apps, and a very large codebase exists (e.g., almost half a million PHP repositories on *GitHub*[20]). The existence of CS in PHP has been confirmed by other researchers and the Software Engineering community has long agreed that, since they are symptoms of poor design, leading to future problems such as reduced maintainability, we should aim at avoiding them. While the best option would be not to insert them during development by means of detection mechanisms embedded in IDEs, we found strong evidence of the presence of 6 CS in 8 widely used PHP web apps, across many years, using the survival analysis technique.

CS occurrences may be removed by explicit refactoring actions or, much less frequently, due to code dropout. It is important for PHP project managers to have an evolutionary perspective on the survival of CS, to decide on the allocation of resources to mitigate their technical debt effects. Since CS are of different types, namely regarding their scope, project managers must be aware of their evolution, with a special concern for scattered CS since their spreads may cause more harm and may be harder to refactor without appropriate tooling. In this study, we found that the survival of localized code smells is around 4 years, while the scattered ones live around 5 years. Around 60% of the CS are removed, and some others, especially scattered ones, are never removed. We also found that the insertion and removal rates are much lower for the scattered CS, than for localized ones.

---

[20] - https://github.com/search?l=PHP&q=PHP&type=Repositories

Since CS has been a recurrent topic taught at universities over, and discussed in practitioners over the last two decades, in theory one would expect that practitioners' attitudes towards CS removal in web apps would have changed throughout time, thereby leading to shorter survival times and less CS density in web apps in more recent years. However, while most of the applications agree with this claim, a consistent trend in CS survival is not observed across all apps.

Therefore, we can classify the development and evolution of the applications into two groups: CS aware and not aware. The increasing complexity of application code, specialty in the number of classes, ads as a strong influence factor to the non-removal of CS. We may consider this as an inertia factor since refactoring a larger and more complex app implies a bigger investment and/or larger schedule. Other factors that influence the results is the number of contributors (team size), and consequently the LOC or LLOC per contributor, and the percentage of non-object-oriented code, especially in the applications with the largest timespan.

Last, but not least, we observed sudden variations in CS occurrence in specific versions, whose root causes deserve investigation and are important for project managers to understand, especially for long-lived projects where managers' turnover inevitably happens from time to time. In this paper we proposed a normalized technique for detecting those anomalies in specific releases/versions during the evolution of web apps, allowing us to unveil the CS story of a development project and make managers aware of the need for enforcing regular refactoring practices. This technique can also be useful in a continuous development pipeline, to add in the quality certification of the release.

Regarding future work, we would like to increase the number of applications and the number of CS studied, provided more computing power is available since collecting CS across many versions is a computationally heavy task. For each application across our long observation periods, we collected CS on the units of a million. It is also worth researching if the results for longitudinal studies on CS depend on the programming language. The obvious way forward here is comparing PHP to Java since many more longitudinal studies on CS exist for the latter.

As a last thought, we believe it makes sense to extend the *technical debt* metaphor (Cunningham, 1992), by introducing the concept of "*CS technical balance of payments*". On the negative side of the balance, we have the *technical debt* caused by code smells insertion. On the positive side of the balance, we have the *technical credit* acquired by investing in refactoring. Getting rid of all CS is probably an unjustifiable quest, since some occurrences may make sense, depending on the technical context. Therefore, the CS technical balance of payments may be slightly negative, just as many countries live well with a negative balance of payments, if it is kept under control.

## Acknowledgments

We are grateful to the late Professor Rui Menezes (deceased 14 May 2019), whose contribution to this work was of great significance. He encouraged and supported us on the usage of survival analysis techniques and inspired us with his enthusiasm.

This work was partially supported by the Portuguese Foundation for Science and Technology (FCT) projects UIDB/04466/2020 e UIDP/04466/2020.